\documentclass[12pt]{iopart}
\usepackage{graphicx} 

\usepackage{subfigure}
 
\begin{document}

\title[Thin film flow and heat transfer over an unsteady stretching sheet]{Thin film flow and heat transfer over an unsteady stretching sheet with thermal radiation, internal heating in presence of external magnetic field}

\author{Prashant G Metri$^{1,2}$, Jagdish Tawade$^3$, \& M Subhas Abel$^2$}

\address{$^1$Division of Applied Mathematics, UKK, M\"{a}lardalen University, V\"{a}ster\aa{}s, Sweden}

\address{$^2$Department of Mathematics Gulbarga University, Gulbarga, Karnataka, India}
\address{$^3$Department of Mathematics, Bheemanna Khandre Institute of Technology Bhalki,Karnataka, India}
\ead{prashant.g.metri@mdh.se, jagadishmaths22@gmail.com}
\vspace{10pt}
\begin{indented}
\item[]
\end{indented}
\begin{abstract}
In this paper we present a mathematical analysis of thin film flow and heat transfer to a laminar liquid film from a horizontal stretching sheet. The flow of thin liquid film and subsequent heat transfer from the stretching surface is investigated with the aid of similarity transformations. Similarity transfor-mations are used to convert unsteady boundary layer equations to a system of non-linear ordinary differential equations. The resulting non-linear differential equations are solved numerically using Runge-kutta-Fehlberg and Newton-Raphson schemes. A relationship between film thickness $\beta$  and the unsteadiness parameter $S$ is found, the effect of unsteadiness parameter $S$, and the  Prandtl number $Pr$, Magnetic field parameter $Mn$, Radiation parameter $Nr$ and viscous dissipation parameter $Ec$ and heat source parameter $\gamma$ on  the temperature distributions are presented and discussed in detail. Present analysis shows that the combined effect of magnetic field, thermal radiation, heat source and viscous dissipation. The  results which form special case of the present study are in excellent agreement with the results reported in the literature.
\end{abstract}

\vspace{2pc}

%\maketitle
\begin{table}[htbp]
\caption{\label{tabone}Nomenclature} 
\begin{indented}
\lineup
\item[]\begin{tabular}{@{}*{4}{l}}
  & & \m & \cr 
\mr
$b$ & Stretching sheet$[s^{-1}]$ &$U$ & Sheet velocity\cr 
& & &  $[m~s^{-1}]$ \cr
$x$ & horizontal coordinate(m) & $y$ & Tvertical coordinate(m)  \cr
$u$ & horizontal velocity component  & $v$ & vertical velocity\cr 
& $[m~s^{-1}]$& & component $[m~s^{-1}]$ \cr
$T$ & Temperature [k] & $t$ & time [s]  \cr
$h$ & film thickness(m) & $S$ & unsteadiness parameter,$\frac{\alpha}{b}$   \cr
$C_p$ & specific heat)$[J kg^{-1} K^{-1}]$ & $f$ & dimensionless streamfunction,\cr 
& & & Eq.(14) \cr 
$Pr$ & Prandtl number,$\frac{\nu}{k}$ & $Ec$ & Eckert number,$\frac{U^2}{C_p(T_s-T_0)} $ \cr
$Mn$ & Magnetic parameter,&  $q$ & heat flux,$-k\frac{\partial T}{\partial y}[Js^{-1} m^{-2}]$ \cr
$Re_x$ & local Reynolds number, & $Nu_{x}$ & local Nusselt number Eq.(27)\cr
$C_f$ & skin friction &  &  \cr
\br
Greek Symbol & & \m & \cr
\mr
$\alpha$ &constant $[s^{-1}]$ & $\eta$ & Similarity variable,\cr 
& & & Eqn.(16) \cr
$\theta$ & Dimensionless temperature Eq.(15) & $\gamma$ &  temperature dependent parameter,\cr 
& & & $\gamma=\frac{Q}{\rho c_{p}b}$  \cr
k & thermal diffusivity $[m^{2} s^{-1}]$ & $\mu$ & dynamic viscosity $[kg m^{-1} s^{-1}]$  \cr
$\nu$ &kinematic viscosity  $[m^{2} s^{-1}]$  & $\rho$ &  density $[kg m^{-3}]$  \cr
$\tau$ & shear stress $\mu\frac{\partial u}{\partial y}[kg m^{-1} s^{-2}]$& $\psi$ & stream function $[m^{2}s^{-1}]$  \cr
\br
Subscripts & & \m & \cr
\mr
o & origin & ref & reference value\cr
s &sheet & x & local value\cr
\br
\end{tabular}
\end{indented}
\end{table}
\section{Introduction}
The analysis of fluid flow and heat transfer across a thin liquid film has attracted the attention of a number of researchers because of its possible practical applications in many branches of science and technology. The knowledge of flow and heat transfer within a thin liquid film is crucial in understanding the coating process and design of various heat exchangers and chemical processing equipment. Other applications include wire and fiber coating, food stuff processing reactor fluidization, transpiration cooling and so on. The prime aim in almost every extrusion application is to maintain the surface quality of the extrudate. All coating processes demand a smooth glossy surface to meet the requirements for best appearance and optimum service properties such as low friction, transparency and strength.The problem of extrusion of thin surface layers needs special attention to gain some knowledge for controlling the coating product efficiently. \\

Heat transfer analysis due to a continuously moving stretching surface through an ambient fluid is one of the thrust areas of current research. This finds its application over a broad spectrum of science and engineering disciplines, especially in the field of chemical engineering. Many chemical engineering processes like metallurgical process, polymer extrusion process involves cooling of a molten liquid being stretched into a cooling system. In such processes the fluid mechanical properties of the penultimate product would mainly depend on two things, one is the cooling liquid used and other is the rate of stretching. For example, in a melt spinning process, the extrudate from the die is generally drawn and simultaneously stretched into a filament or sheet, which is thereafter solidified through rapid quenching or gradual cooling by direct contact with water or the coolant liquid. The stretching provides a unidirectional orientation to the extrudate, thereby improving the fluid mechanical properties.  The quality of the final product greatly depends on the rate of cooling and the stretching rate.  The choice of an appropriate cooling liquid is crucial as it has a direct impact on rate of cooling and care must be taken to exercise optimum stretching rate otherwise sudden stretching may spoil the properties desired for the final outcome.  Some industrially important liquids like synthetic oils, dilute polymeric solutions such as 5.4\% of polyisobutylene in cetane can be used as effective coolant liquids (see~\cite{1}). Since the quality of the final product in such extrusion processes depend considerably on the flow and heat transfer characteristics within a thin liquid film over a stretching sheet, the analysis and fundamental understanding of the momentum and thermal transports for such processes are very important.\\

In 1961, Sakiadis ~\cite{2} initiated the study of boundary layer flow over a continuous solid surface with constant speed. This flow was Blasius type, in which the boundary layer thickness is increased with distance from the slit. Crane~\cite{3} was the first among others to consider the steady two-dimensional flow of a Newtonian fluid driven by a stretching elastic flat sheet which moves in its own plane with a velocity varying linearly with the distance from a fixed point. Bujurke~\cite{4} analysed the second order fluid over a stretching sheet using momentum integral technique. Ray et al.~\cite{5} developed the thin liquid film on surface of spinning disk in presence of transverse magnetic field. The film thickness is supress within the spinning time but increase in the magnetic field for fixed time. Dandpat et al.~\cite{6} analysed thin film flow on rotating disc under the action of thermocapillary force. Andersson et al.~\cite{7} studied heat transfer in a liquid film on an horizontal stretching sheet. He also investigated film thickeness and unsteady stretching sheet. Dandpat et al.~\cite{8} analyzed thermocapillarity flow in a liquid film on a horizontal stretching sheet. Thermocapillarity influence to thicken the film and increase the rate of heat transfer between the sheet and film. Wang~\cite{9} analyzed the thin film flow over a horizontal stretching sheet. Analytically solved by homotopy analysis method it is helful to understand the flow and heat transfer mechanisms of the liquid film. Dandpat et al.~\cite{10} studied two diemensional laminar thin film flow on an unsteady stretching sheet. Momentum equations are solved analytically in flow situations by using singular petrubution method. Santra et al.~\cite{11} analyzed the thermocapillary effects on unsteady thin film flow over a heated hoeizontal stretching sheet. The effect of thermocapillary in the temperature distrubution  in stretching direction decreases at higher values of Prandtl number and Biot number. Noor et al.~\cite{12} studied the effects of themocapillary and magnetic field in a thin film flow over an unsteady stretching surface. Analytical method is solved by homotopy analysis method to show the flow and heat transfer rate of the thin liquid film. Dandpat et al.~\cite{13} shown the effects of variable fluid properties on thin liquid film flow over an unsteady heated stretching surface.The effect of variable fluid viscosity on velocity profile  increases when viscosity decreases due to decrease of temperature along the stretching sheet.\\

There are extensive works in literature concerning the production of thin fluid film either on a vertical wall achieved through the action of gravity or that over a rotating disc achieved through the action of centrifugal forces. Dandpat et al.~\cite{14} studied numerically two layer thin film flow on a rotating disk under the assumptions of planar interface and free surface. Numerically solved by finite element method to show the film thickness  varies at different time zone  depend on the rate of ratational speed of the disk. Dandpat et al.~\cite{15} analyzed magneto hydromagnetic thin film flow over an unsteady heated stretching sheet. Film thickness equations are derived using long wave approximation theory of thin liquid film. Li et al ~\cite{16} studied thin film flo and heat transfer over an unsteady stretching sheet in presence of thermal radiation, variable heat flux and internal heating. Thermal radiation leads to increase in temperature distrubution. It means to fall the rate of cooling for the thin film flow. Vajravelu et al~\cite{17} explore the the effects of  variable thermal conductivity and the viscous dissipation on the heat transfer of an incompressible power-law liquid film flow over an unsteady porous stretching sheet. Thermal conductivity, viscous dissipation and  powe law index plays significant role in heat transfer process. Aziz et al.~\cite{18} investigated thin film flow thin film flow and heat transfer over an unsteady permeable stretching sheet. Numerically solved by homotopy analysis method to show  flow and heat transfer rate of the thin liquid film. Gul et al.~\cite{19} analyzed the hydromagnetic thin flim flow over an unsteady second grade fluid past a verticle oscilating belt. Analytically solved by Adomain decomposition method and optimal homotopy asymptotic method. Megahed et al.~\cite{20} studied casson thin film flow over an unsteady stretching sheet with viscous dissipation and variable heat flux involving boundary conditions of slip effects. Viscous dissipation and heat flux  coupled with the velocity slip effects plays significant role in the rate of heat transfer.Alice et al. ~\cite{21} investigated the effects of feedback control, applied via a suction boundary condition and based observation of the interference height on the dynamic and stability of a thin layer of fluid flowing down in inclined plane. Here they con-sidered two different models Benny and weighted residual models that accounts for the fluid injection through the wall. \\
The purpose of present study is to give numerical analysis of hydromagnetic  thin film flow over an unsteady stretching sheet in presence of thermal radiation and interna heating with uniform film thickness. The governing equations are transformed into highly non-linear ordinary differential equations and then solved numerically by using Runge-Kutta-Fehlberg and newton-Raphason schemes based on shooting technique. Numerical computation has been carried out for thermal boundary layer for various values of flow parameters. Comparison with known results (Wang~\cite{9}) for certain particular cases is in excellent agreement.

\section{Formulation of the Problem}
\begin{figure}
\centering
\includegraphics[scale=0.75]{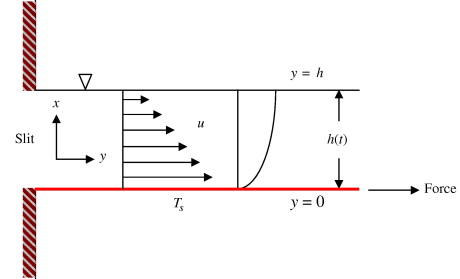}
\caption{Schematic representation of a liquid film on an elastic sheet}
\label{fig1}       
\end{figure}

Let us consider a thin elastic sheet which emerges from a narrow slit at the origin of a Cartesian co-ordinate system for investigations as shown schematically in Fig.~\ref{fig1}. The continuous sheet at $y$ = 0 is parallel with the $x$-axis and moves in its own plane with the velocity.

\begin{equation}
\label{eq1}
\ U (x,t)=\frac{bx}{1-\alpha t},
\end{equation}

where $b$ and $\alpha$  are both positive constants with dimension per time. The surface temperature $T_{s}$  of the stretching sheet is assumed to vary with the distance $x$ from the slit as 
\begin{equation}
\label{eq2}
\ T_{s} (x,t)=T_{0}-T_{ref} \left[\frac{bx^{2}}{2\nu}\right] ({1-\alpha t})^{-\frac{3}{2}},
\end{equation}
where $T_{0}$ is the temperature at the slit and $T_{ref}$ can be taken as a constant reference temperature such that $0 \leq T_{ref}\leq T_0$. The term $\frac{bx^2}{\nu(1-\alpha t)}$ can be recognized as the local Reynolds number based on the surface velocity $U$. The expression (\ref{eq1}) for the velocity
of the sheet $U(x,t)$ reflects that the elastic sheet which is fixed at the origin is stretched by applying a force in the positive
$x$-direction and the effective stretching rate $\frac{b}{(1-\alpha t)}$ increase with time as $0 \leq \alpha<1 $. With the same analogy the expression for the surface temperature $T_s(x,t)$ given by (\ref{eq2}) represents a situation in which the sheet temperature decreases from $T_0$ at the slit in proportion to $x^2$ and such that the amount of temperature reduction along the sheet increases with time. The applied transverse magnetic field is assumed to be of variable kind and is chosen in its special form as
\begin{equation}
 \label{eq3}
\ B (x,t)=B_{0} ({1-\alpha t})^{-\frac{1}{2}}. 
\end{equation}

The particular form of the expressions for $U(x, t)$, $T_s (x,t)$ and $B(x, t)$ are chosen so as to facilitate the construction of a new similarity transformation which enables in transforming the governing partial differential equations of momentum and heat transport into a set of non-linear ordinary differential equations.\\

Consider a thin elastic thin liquid film of uniform thickness $h(t)$ lying on the horizontal stretching sheet (Fig.~\ref{fig1}). The $x$-axis is chosen in the direction along which the sheet is set to motion and the $y$-axis is taken perpendicular to it. The fluid motion within the film is primarily caused solely by stretching of the sheet. The sheet is stretched by the action of two equal and 
opposite forces along the $x$-axis. The sheet is assumed to have velocity $U$ as defined in (\ref{eq1}) and the flow field is exposed to the influence of an external transverse magnetic field of strength $B$ as defined in (\ref{eq3}). We have neglected the effect of latent heat due to evaporation by assuming the liquid to be nonvolatile. Further the buoyancy is neglected due to the relatively thin liquid film, but it is not so thin that intermolecular forces come into play. The velocity and temperature fields of the thin liquid film obey the following boundary layer equations\\

\begin{equation}
\label{eq4}
  \frac{\partial u}{\partial x}
   +   \frac{\partial u}{\partial y}
      =0,
\end{equation}

\begin{equation}
\label{eq5}
\frac{\partial u}{\partial t}+u  \frac{\partial u}{\partial x}+v  \frac{\partial u}{\partial y} =\nu  \frac{\partial^2 u}{\partial  y^2} -\frac{\sigma B^{2}_{0}}{\rho} u,   
\end{equation}

\begin{equation}
\label{eq6}
 \frac{\partial T}{\partial t}+u  \frac{\partial T}{\partial x}+\nu  \frac{\partial T}{\partial y}=\frac{k}{\rho C_{p}} \frac{\partial^2 T}{\partial y^2} +\frac{\mu}{\rho C_p} \left(\frac{\partial u}{\partial y}\right)^2 -\frac{1}{\rho C_{p}}\frac{\partial q_{r}}{\partial y}+\frac{Q}{\rho C_{p}}(T_{s}-T_{0}).
\end{equation}

Following Rosseland approximation (see~\cite{23}) the radiative heat flux $ q_{r} $  and  is modelled as,

\begin{equation}
\label{eq7}
 \frac{\partial q_{r}}{\partial y}=-\frac{4 \sigma^*}{3 k^*} \frac{\partial (T^4)}{\partial y},  
\end{equation}

where $ \sigma^* $  is the Stefan-Boltzmann constant and $ k^* $  is the mean absorption coefficient. Assuming that the differences in temperature within the flow are such that $ T^4 $  can be expressed as a linear combination of the temperature, we expand  $ T^4 $   in a Taylor’s series about  $ T_\infty $ as follows

\begin{equation*}
T^4=T_{\infty}^{4}+4T^{3}_{\infty}(T-T_\infty)+6T_{\infty}^{2} (T-T_\infty)^{2} +....,
\end{equation*}
and neglecting higher order terms beyond the first degree in $ (T-T_\infty) $  we get

\begin{equation}
\label{eq8}
  T^4\cong -3 T_{\infty}^{4}+ 4T^{3}_{\infty} T.
\end{equation}

Substituting (\ref{eq8}) in (\ref{eq7}) we obtain
\begin{equation}
\label{eq9}
  \frac{\partial q_{r}}{\partial y}=-\frac{16T_{\infty}^{*}\sigma^{*}}{3k^{*}} \frac{\partial^{2}T^{4}}{\partial y^{2}}.
\end{equation}

Using (\ref{eq9}) in (\ref{eq6}) we obtain 
\begin{eqnarray}
\label{eq10}
 \frac{\partial T}{\partial t}+u  \frac{\partial T}{\partial x}+\nu  \frac{\partial T}{\partial y}= \nonumber \\
=\frac{k}{\rho C_{p}} \frac{\partial^2 T}{\partial y^2} +\frac{\mu}{\rho C_p} \left(\frac{\partial u}{\partial y}\right)^2-\frac{1}{\rho C_{p}}\left(-\frac{16T_{\infty}^{*}\sigma^{*}}{3k^{*}} \frac{\partial^{2}T^{4}}{\partial y^{2}}\right)+\frac{Q}{\rho C_{P}}(T_{s}-T_{0}).
\end{eqnarray}

The pressure in the surrounding gas phase is assumed to be uniform and the gravity force gives rise to a hydrostatic pressure variation in the liquid film. In order to justify the boundary layer approximation, the length scale in the primary flow direction must be significantly larger than the length scale in the cross stream direction. We choose the representative measure of the film thickness to be $(\frac{\nu}{b}) ^{\frac{1}{2}}$ so that the scale ratio is large enough i.e. $\frac{x}{(\frac{\nu}{b}) ^{\frac{1}{2}}}\gg$. This choice of length scale enables us to
employ the boundary layer approximations. Further it is assumed that the induced Magnetic field is negligibly small.

The associated boundary conditions are given by
\begin{equation}
\label{eq11}
u=U,~~~\nu=0,~~~T=T_s~~~\mathrm{at}~~~y=0,
 \end{equation}
 
 \begin{equation}
 \label{eq12}
 \frac{\partial u}{\partial y}=\frac{\partial T}{\partial y}=0~~~\mathrm{at}~~~y=h,
\end{equation}

 \begin{equation}
 \label{eq13}
\nu=\frac{dh}{dt}~~~\mathrm{at}~~~y=h.
\end{equation}

At this juncture we make a note that the mathematical problem is implicitly formulated only for $x\leq0$. Further it is assumed that the surface of the planar liquid film is smooth so as to avoid the complications due to surface waves. The influence
of interfacial shear due to the quiescent atmosphere, in other words the effect of surface tension is assumed to be negligible. The viscous shear stress $\tau=\mu(\frac{\partial u}{\partial y})$
and the heat flux $q=-k(\frac{\partial T}{\partial y})$ vanish at the adiabatic free surface (at $y = h$).

\section{Similarity Transformations}
We introduce diemensionaless variables $f$ and $\theta$ and the similarity variable $\eta$ as

 \begin{equation}
 \label{eq14}
f(\eta)=\frac{\psi (x,y,t)}{(\frac{\nu b}{1-\alpha t})^{\frac{1}{2}}},
 \end{equation}

 \begin{equation}
 \label{eq15}
\theta (\eta)=\frac{T_{0}-T(x,y,t)}{T_{ref}\left(\frac{bx^{2}}{2\nu(1-\alpha t)^{-\frac{3}{2}}}\right)},
 \end{equation}
  
\begin{equation}
\label{eq16}
\eta=(\frac{b}{\nu(1-\alpha t)})^{\frac{1}{2}} y.
 \end{equation}

The physical stream function $\psi(x, y, t)$ automatically assures mass conservation given in (\ref{eq4}). The velocity components
are readily obtained as

 \begin{equation}
 \label{eq17}
u=\frac{\partial \psi}{\partial y}=\left(\frac{bx}{1-\alpha t}\right)f^{'}(\eta),
 \end{equation}
 
 \begin{equation}
 \label{eq18}
 \nu=-\frac{\partial \psi}{\partial x}=-\left(\frac{\nu b}{1-\alpha t}\right)^{\frac{1}{2}}f(\eta).
 \end{equation}
 
 The mathematical problem defined through (\ref{eq5})-(\ref{eq6}) and (\ref{eq11})-(\ref{eq13}) transforms exactly into a set of ordinary differential equations
and associated boundary conditions as follows

\begin{equation}
\label{eq19}
 S\left(f^{'}+\frac{\eta}{2}f^{''}\right)+(f^{'})^{2}-ff^{''}=f^{'''}-Mnf^{'},
 \end{equation}
 
 \begin{eqnarray}
 \label{eq20}
Pr \left[\frac{S}{2}(3\theta + \eta \theta^{'})+(2f^{'}-\gamma)\theta-\theta^{'}f\right]=\nonumber \\ \theta^{''}(1+Nr)-Ec Pr f''^{2},
\end{eqnarray}

 \begin{equation}
 \label{eq21}
f^{'}(0)=1,~~~f(0)=0,~~~\theta(0)=1, 
 \end{equation}
 
  \begin{equation}
  \label{eq22}
 f^{''}(\beta)=0,~~~\theta^{'}(\beta)=0, 
 \end{equation}
 
  \begin{equation}
  \label{eq23}
  f(\beta)=\frac{S \beta}{2}.
 \end{equation}

Here $S=\frac{\alpha}{b}$
is the dimensionless measure of the unsteadiness and the prime indicates differentiation with respect to $\eta$.
Further, $\beta$ denotes the value of the similarity variable $\eta$ at the free surface so that (\ref{eq16}) gives

 \begin{equation}
 \label{eq24}
 \beta=\left(\frac{b}{\nu(1-\alpha t)}\right)^{\frac{1}{2}}h.
 \end{equation}
 
Yet $\beta$ is an unknown constant, which should be determined as an integral part of the boundary value problem. The rate at
which film thickness varies can be obtained by differentiating (\ref{eq24}) with respect to $t$, in the form

 \begin{equation}
 \label{eq25}
\frac{dh}{dt}=-\frac{\alpha \beta}{2} \left(\frac{\nu}{b(1-\alpha t)}\right)^{\frac{1}{2}}.
 \end{equation}
 
Thus the kinematic constraint at $y = h(t)$ given by (\ref{eq13}) transforms into the free surface condition (\ref{eq25}). It is noteworthy
that the momentum boundary layer equation defined by (\ref{eq19}) subject to the relevant boundary conditions (\ref{eq21})-(\ref{eq23}) is
decoupled from the thermal field, on the other hand the temperature field $\theta(\eta)$ is coupled with the velocity field $f(\eta)$. Since the sheet is stretched horizontally the convection least affects the flow (i.e., buoyancy effect is negligibly small) and hence there is a one-way coupling of velocity and thermal fields.

The local skin friction coefficient, which of practical importance, is given by

\begin{equation}
\label{eq26}
 C_{f}=\frac{-2\mu(\frac{\partial u}{\partial y})_{y=0}}{\rho U^{2}}=-2Re^{-\frac{1}{2}}_{x} f^{''}(0),
 \end{equation} 
 
\begin{equation}
\label{eq27}
Nu_{x}=-\frac{x}{T_{ref}} \left(\frac{\partial T}{\partial y}\right)_{y=0}=\frac{1}{2}(1-\alpha t)^{\frac{-1}{2}} \theta^{'}(0)Re^{\frac{3}{2}}_{x},
 \end{equation}
where $Re_{x}=\frac{Ux}{\nu}$
t the local Reynolds number and $T_{ref}$ denotes the same reference temperature (temperature difference) as in (\ref{eq2}).

We now march on to find the solution of the boundary value problem (\ref{eq19})-(\ref{eq23}).
  
 \section{Numerical Solution}
The system of non-linear differential equations (\ref{eq19}) and (\ref{eq20}) subjected to the  boundary conditions (\ref{eq21})-(\ref{eq23}) are solved numerically, using  Runge-Kutta-Fehlberg and Newton-Raphson schemes based shooting method.   (see~\cite{22}). In this method, third order non-linear  ordinary differential equation (\ref{eq19}) and second order non-linear  ordinary differential equation (\ref{eq20}) have been reduced to  first order differential equations as follows:

\begin{equation}
\label{eq28}
\frac{df_{0}}{d\eta}=f_{1}, 
 \end{equation}

\begin{equation}
\label{eq29}
\frac{df_{1}}{d\eta}=f_{2},
 \end{equation}
 
 \begin{equation}
 \label{eq30}
\frac{df_{2}}{d\eta}=S\left(f_{1}+\frac{\eta}{2}f_{2}\right)+(f_{1})^{2}-f_{0}f_{2}+Mnf_{1},
 \end{equation}

\begin{equation}
\label{eq31}
\frac{d\theta_{0}}{d\eta}=\theta_{1},
 \end{equation}
 
 \begin{equation}
 \label{eq32}
 \frac{d\theta_{1}}{d\eta}=\biggl(Pr \left[\frac{S}{2}(3\theta_{0} + \eta \theta_{1})+(2f_{1}-\gamma)\theta_{0}-\theta_{1}f_{0}\right] 
-Ec Pr f_{2}^{2}\biggr).
 \end{equation}

 Corresponding boundary conditions take the form,
 \begin{equation}
 \label{eq33}
f_{1}(0)=1,~~~f_{0}(0)=0,~~~\theta_{0}(0)=1, 
 \end{equation}
 
   \begin{equation}
   \label{eq34}
f_{2}(\beta)=0,~~~\theta_{1}(\beta)=0, 
 \end{equation}
 
  \begin{equation}
  \label{eq35}
 f_{0}(\beta)=\frac{S \beta}{2}.
 \end{equation}
 
  Here $f_{0}(\eta)=f(\eta)$ and $\theta_{0}(\eta)=\theta(\eta)$. The above boundary value problem is first converted into an initial value problem by appropriately guessing the missing slopes $f_{2}(0)$ and $\theta_{1}(0)$. The resulting IVP is solved by shooting method for a set of parameters appearing in the governing equations and a known value of $S$. The value of $\beta$ is so adjusted that condition (\ref{eq35}) holds. This is done on the trial and error basis. The value for which condition (\ref{eq35}) holds is taken as the appropriate film thickness and the IVP is finally solved using this value of $\beta$. The step length of $h = 0.01$ is employed for the computation purpose. The convergence criterion largely depends on fairly good guesses of the initial conditions in the shooting technique. The iterative process is terminated until the relative difference between the current and the previous iterative values of $f(\beta)$ matches with the value of $\frac{S\beta}{2}$ up to a tolerance of $10^{-6}$. Once the convergence in achieved we employed shooting technique with the Runge-Kutta-Fehlberg and Newton-Raphason schemes to determine the unknown in order to convert the boundary value problem to initial value problem. Once all initial conditions are determined, the resulting differential equations are determined, the resulting differential equations were integrated using initial value solver. For this purpose Runge-Kutta-Fehlberg scheme was used.

 \section{Results and Disscussion}
\begin{figure}[htbp]
\centering
\includegraphics[width=.55\textwidth]{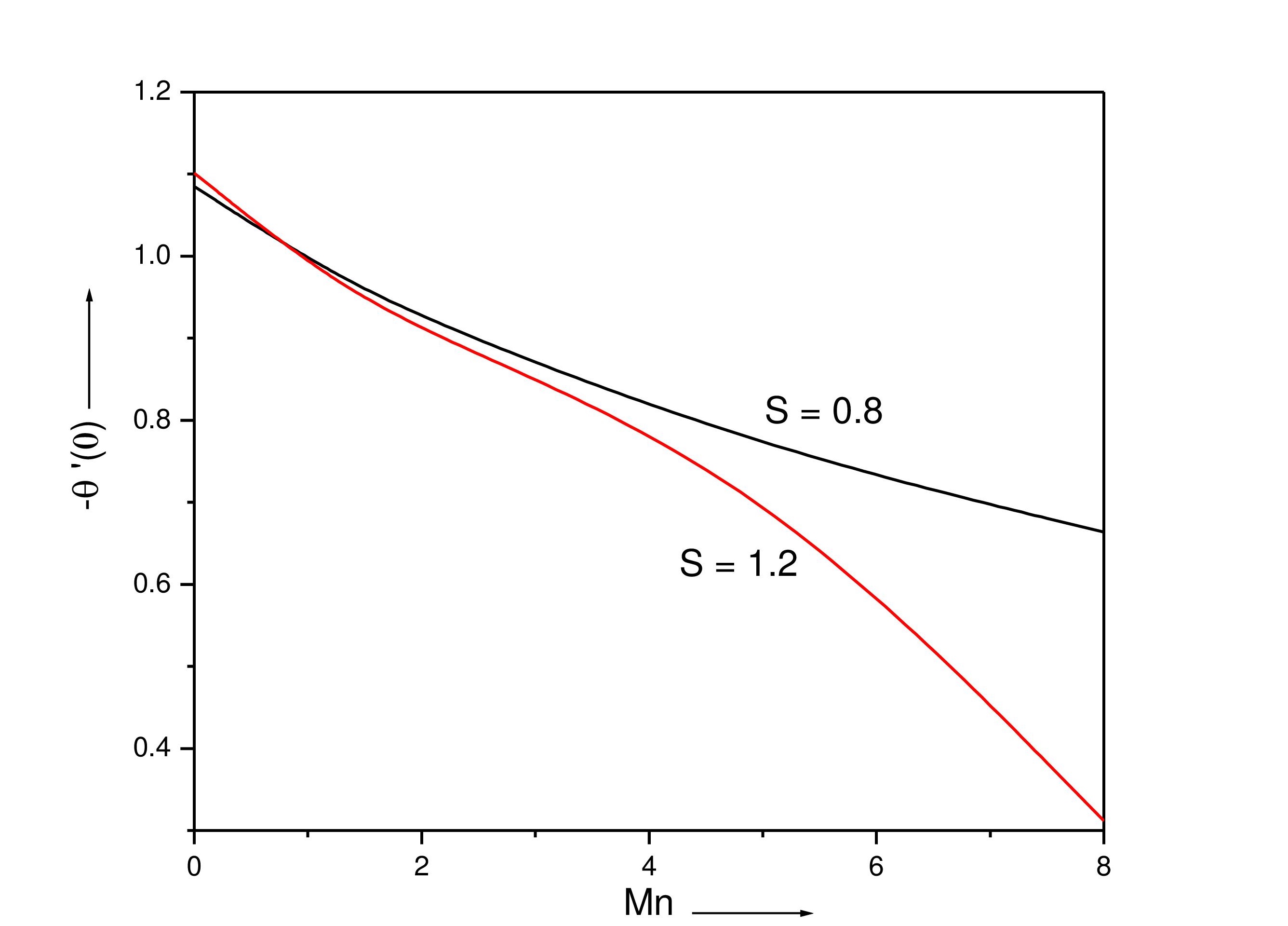}
\caption{Diemensionless temperature gradiant $ -\theta'(\eta) $ at the sheet vs magnetic parameter $Mn$ for $S$=0.8 and $S$=1.2}
\label{fig2}       
\end{figure}

\begin{figure}[htbp]
  \centering
 \subfigure[S=0.8]{\includegraphics[height=5.6cm,keepaspectratio]{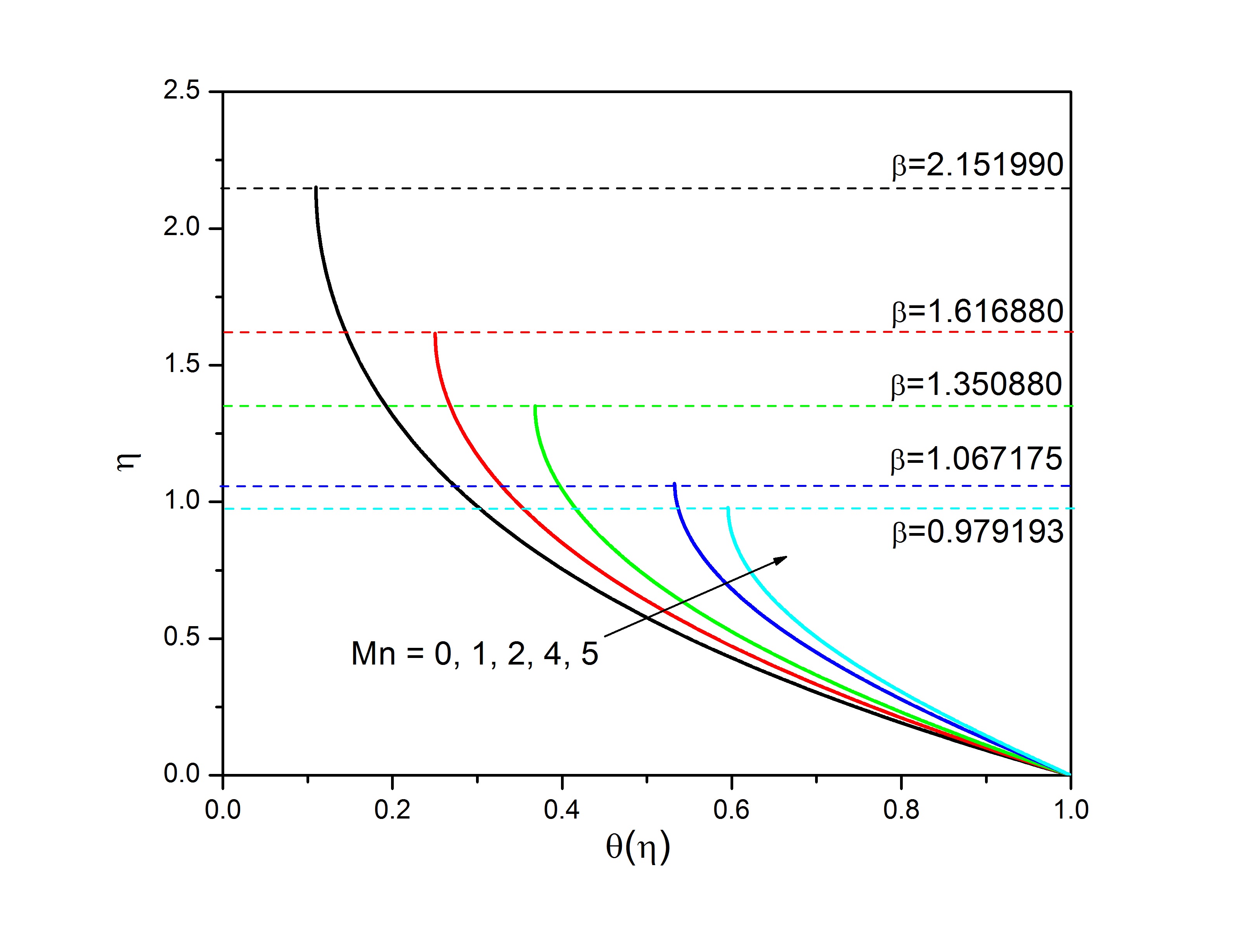}}
\subfigure[S=1.2]{\includegraphics[height=5.6cm,keepaspectratio]{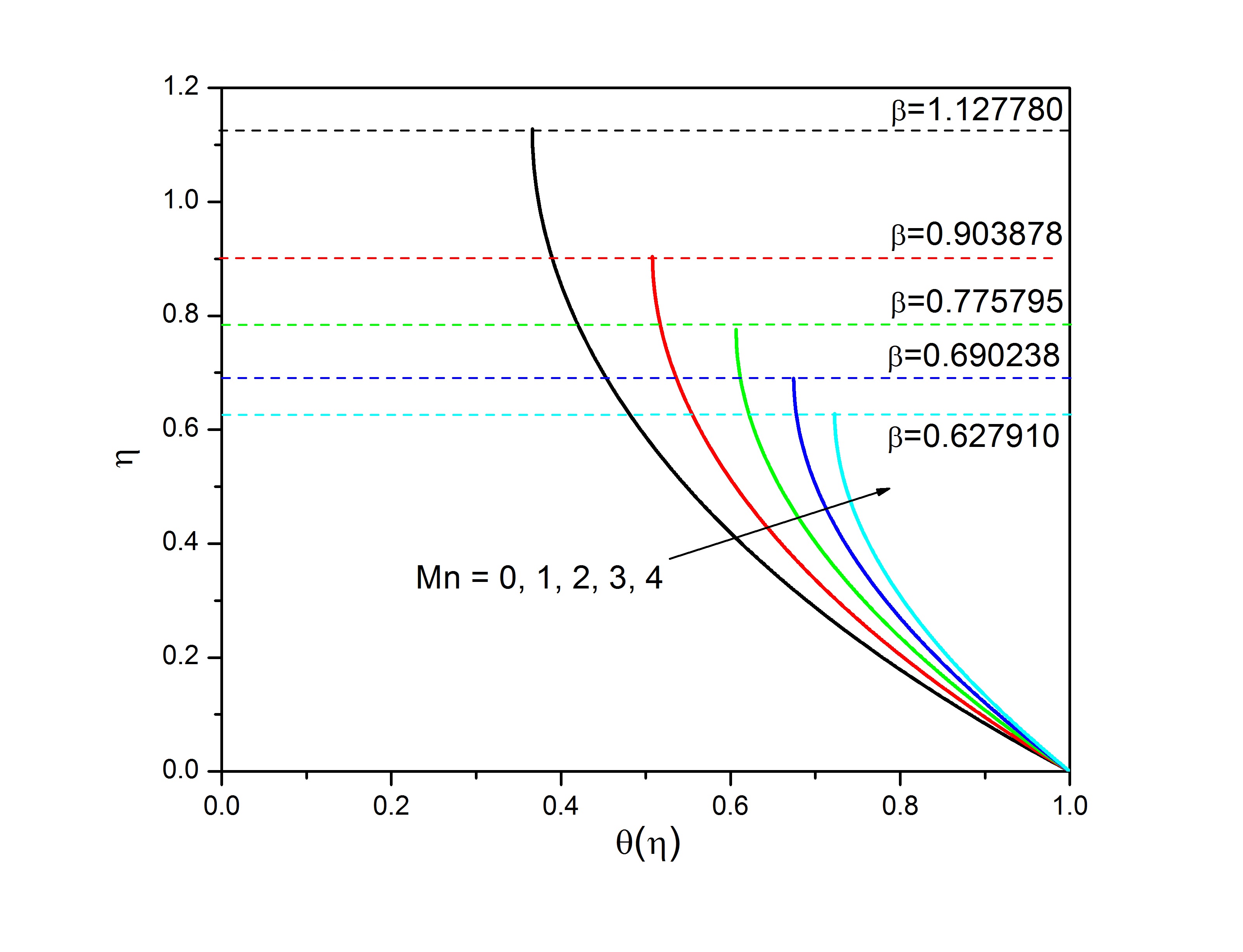}}
\caption{Variation in the temperature profile $ \theta(\eta) $ for different values of Magnetic parameter $Mn$}
 \label{fig3}
\end{figure}

\begin{figure}[htbp]
  \centering
\subfigure[S=0.8]{\includegraphics[height=5cm,keepaspectratio]{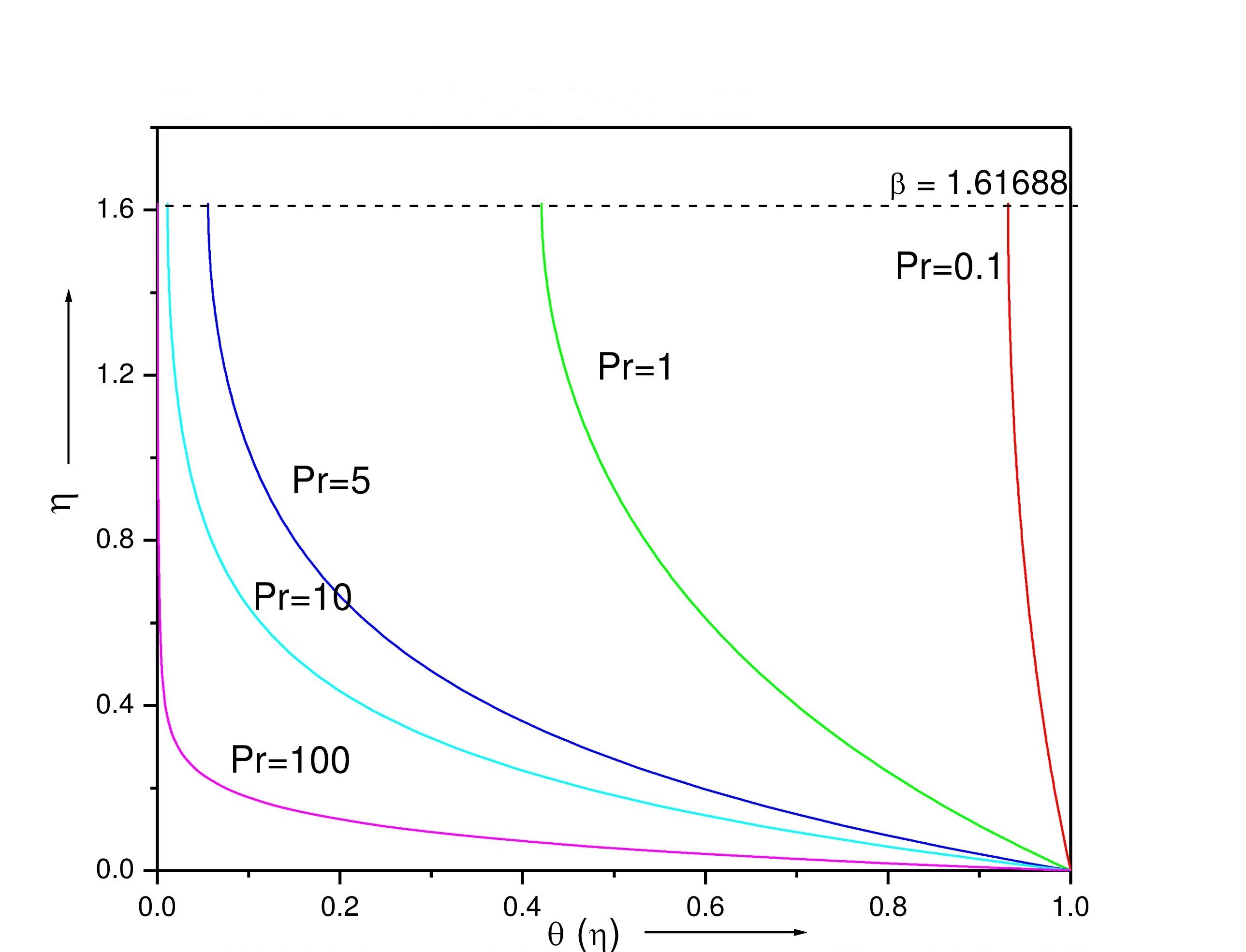}}
\subfigure[S=1.2]{\includegraphics[height=5cm,keepaspectratio]{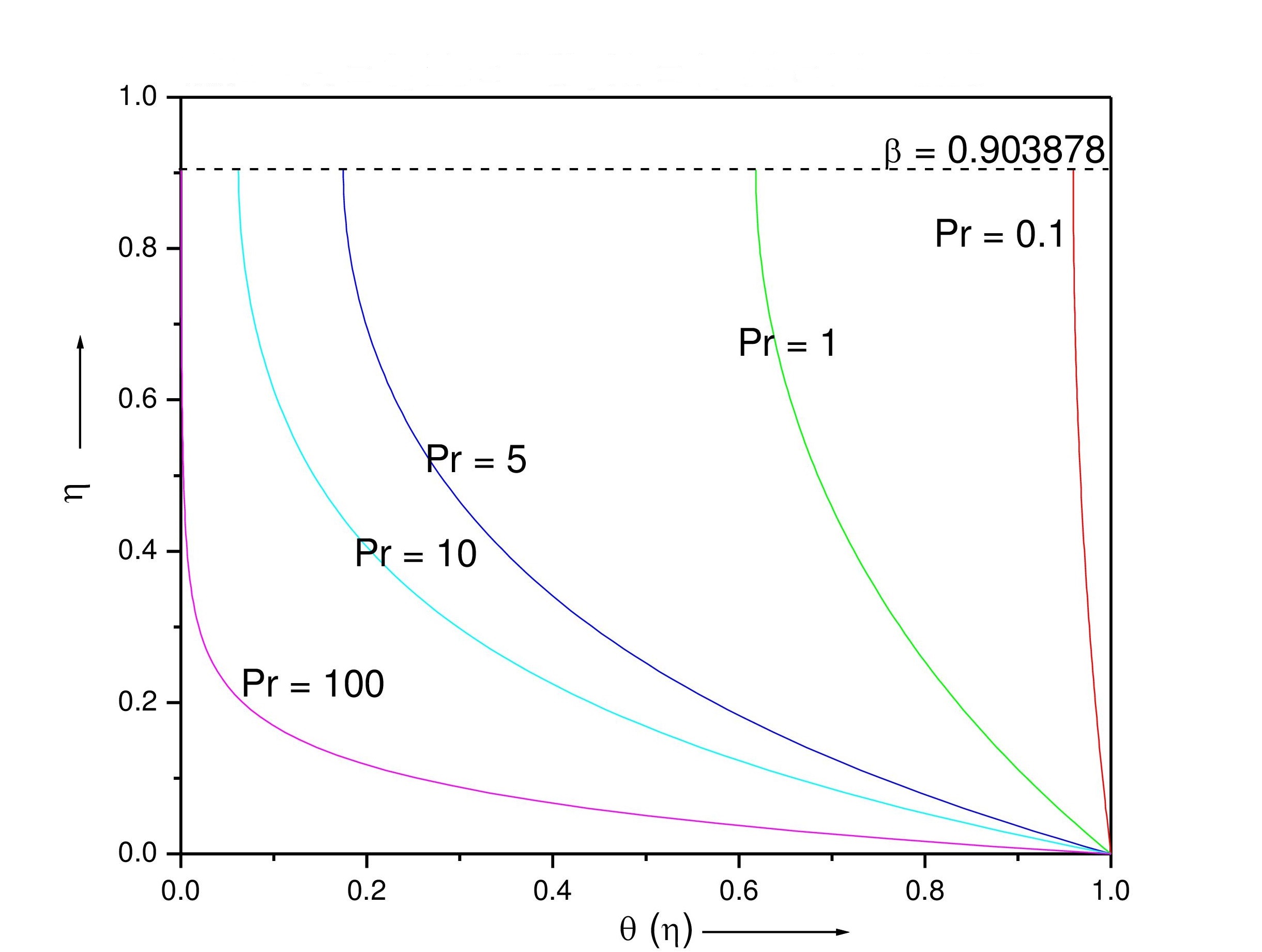}}
\caption{Variation in the temperature profile $ \theta(\eta) $ for different values of Prandtl number $Pr$}
 \label{fig4}
\end{figure}

\begin{figure}[htbp]
  \centering  
\subfigure[S=0.8]{\includegraphics[height=5cm,keepaspectratio]{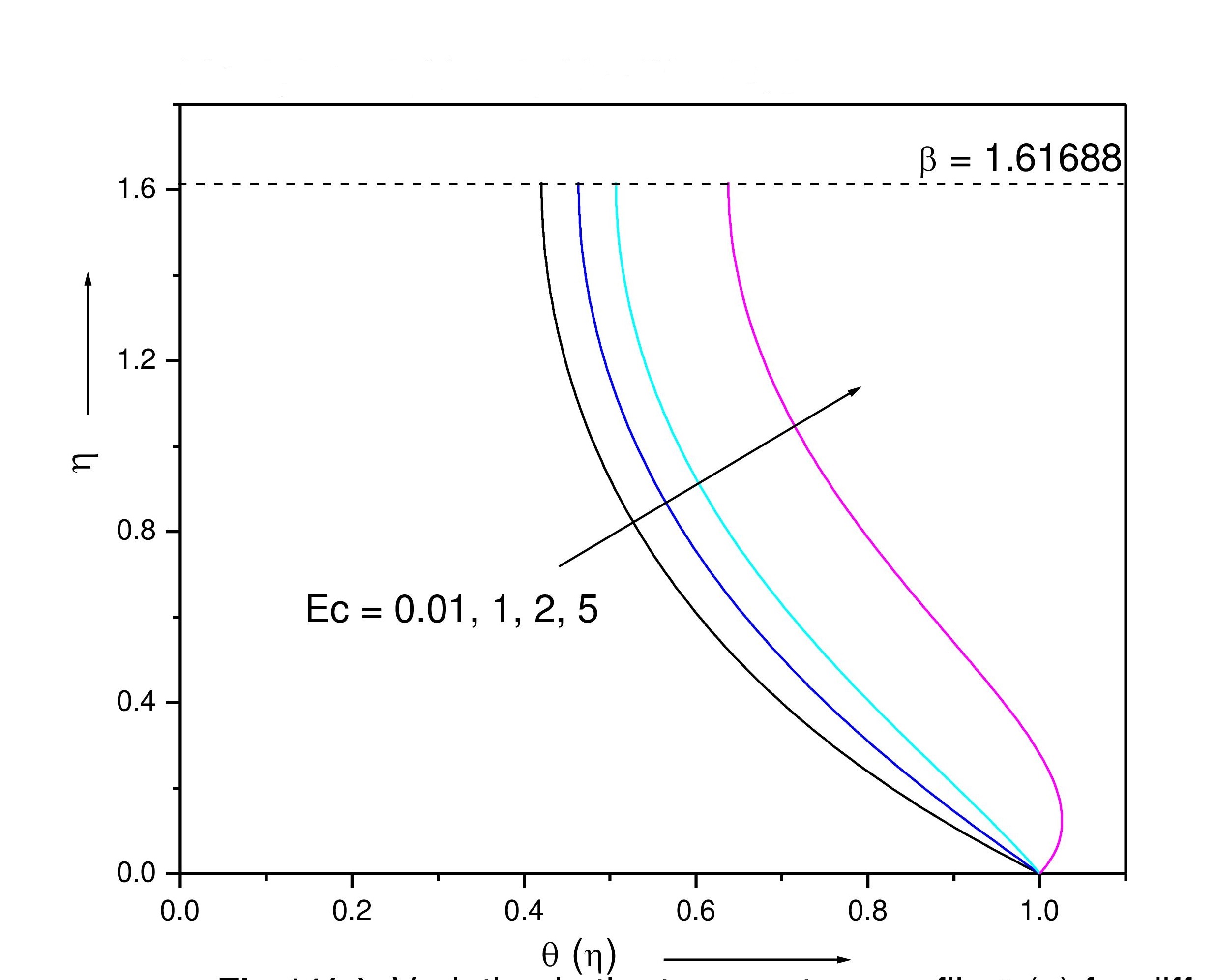}}
\subfigure[S=1.2]{\includegraphics[height=5cm,keepaspectratio]{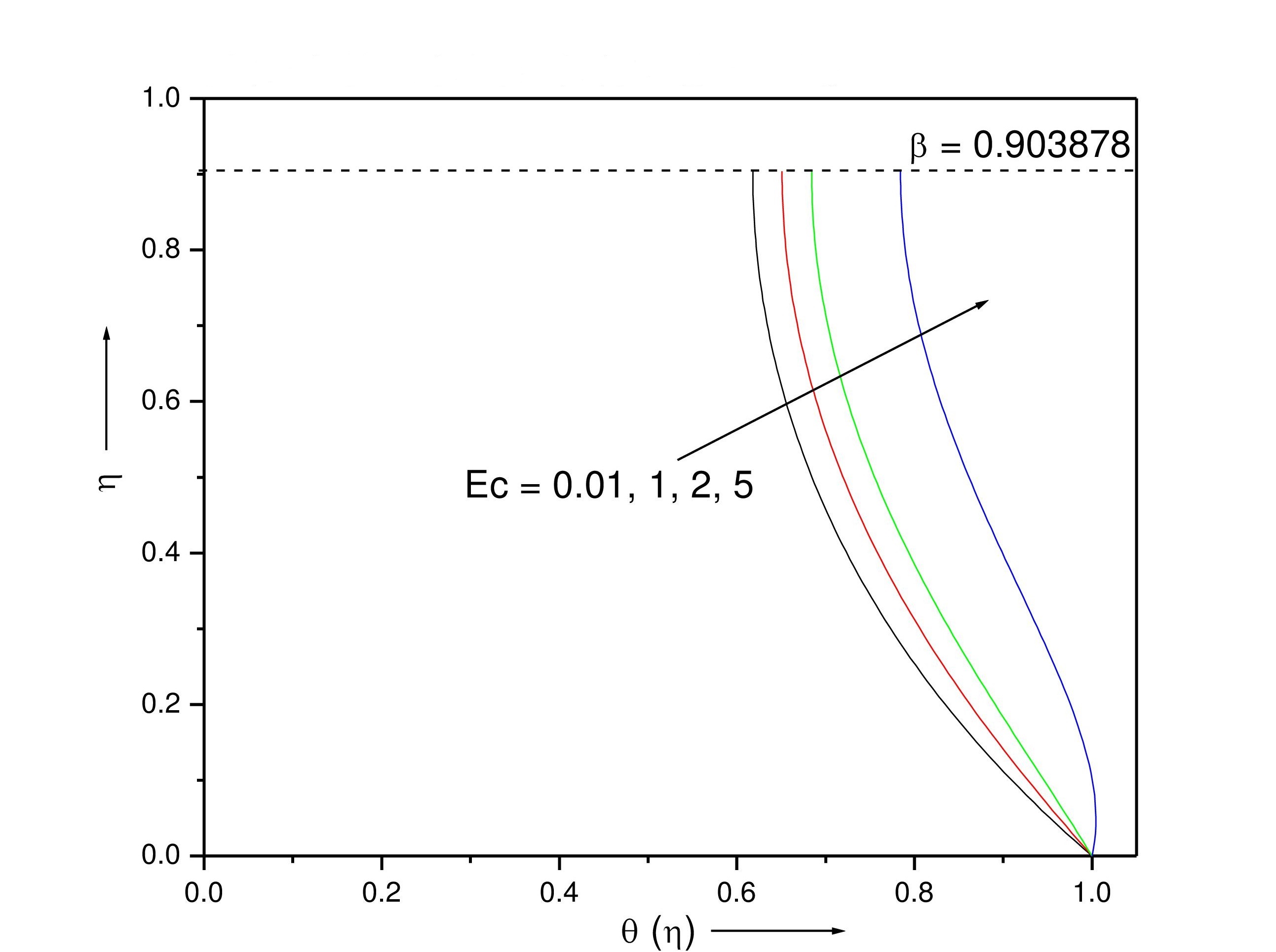}}
\caption{Variation in the temperature profile $ \theta(\eta) $ for different values of Eckert number $Ec$}
\label{fig5}
\end{figure}

\begin{figure}[htbp]
  \centering 
\subfigure[S=0.8]{\includegraphics[height=5cm,keepaspectratio]{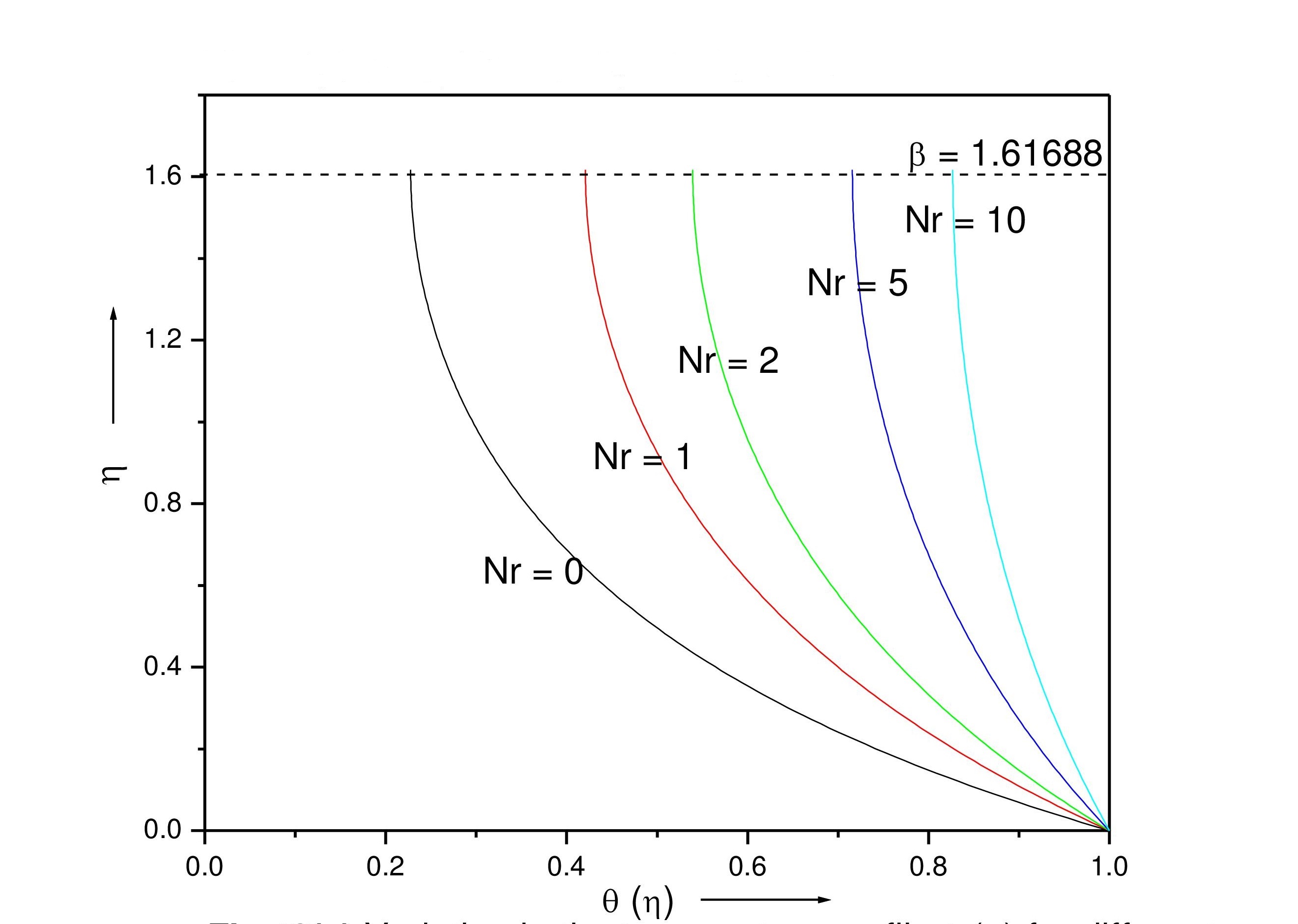}}
\subfigure[S=1.2]{\includegraphics[height=5cm,keepaspectratio]{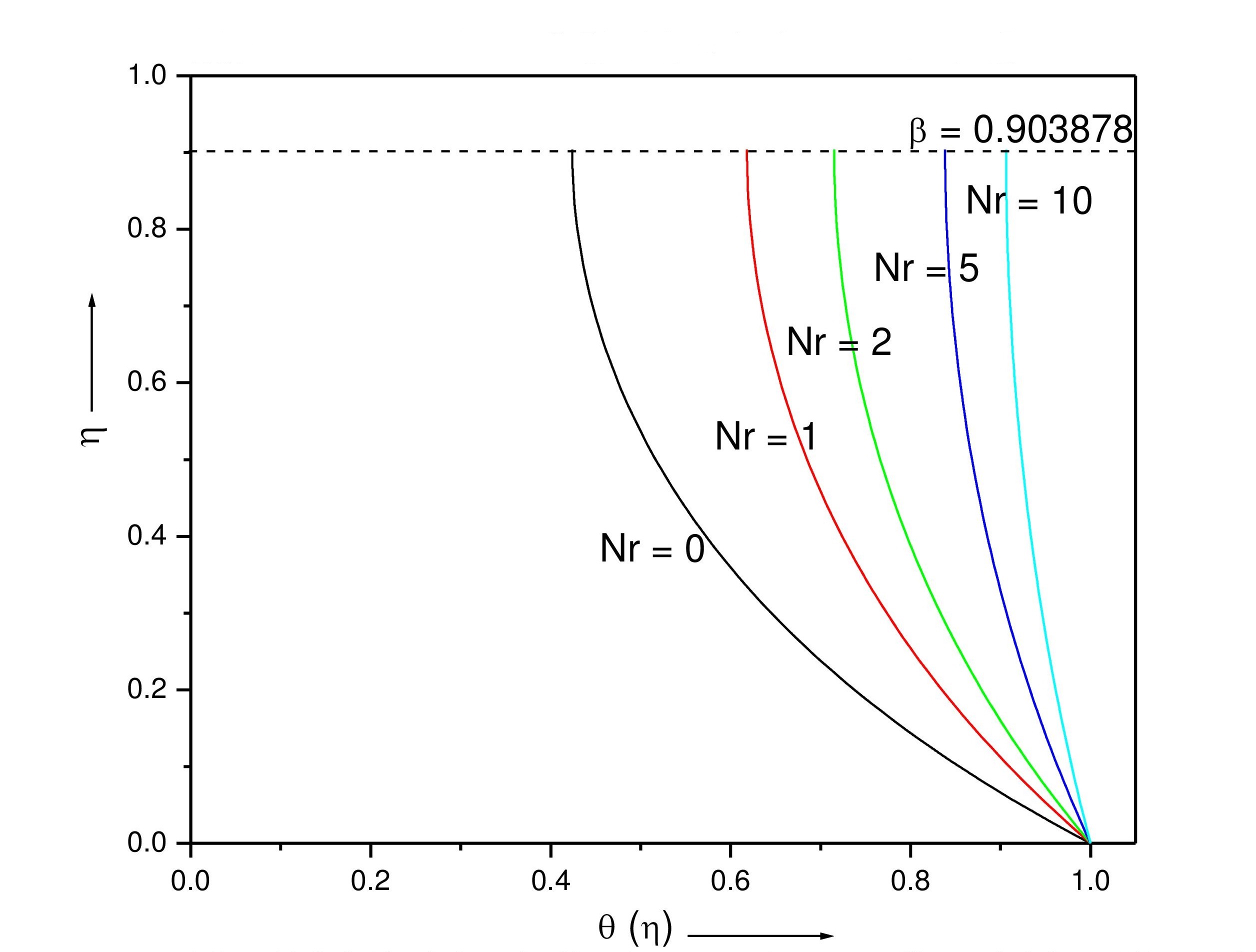}}
\caption{Variation in the temperature profile $ \theta(\eta) $ for different values of Radiation parameter $Nr$}
 \label{fig6}
\end{figure}

\begin{figure}[htbp]
  \centering 
\subfigure[S=0.8]{\includegraphics[height=5.6cm,keepaspectratio]{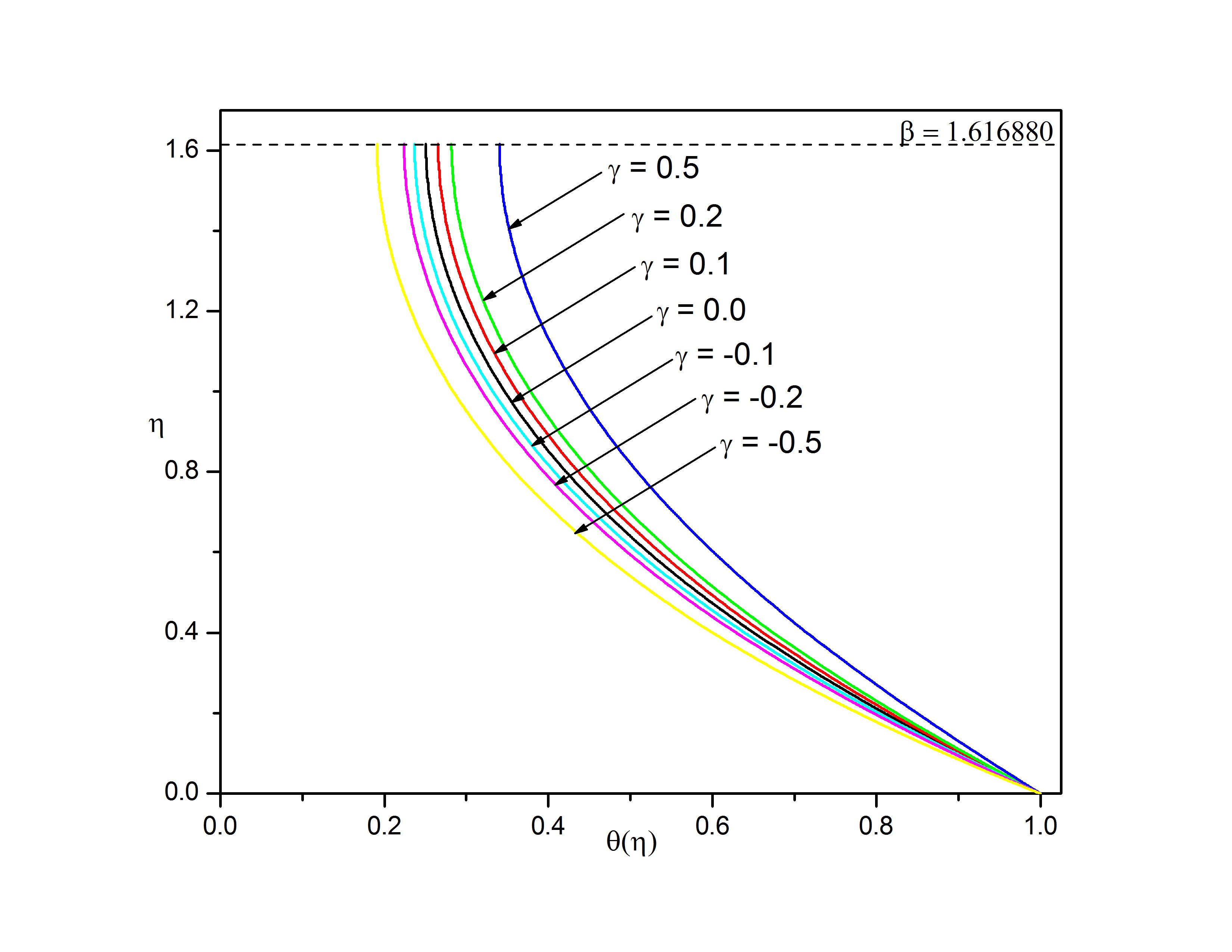}}
\subfigure[S=1.2]{\includegraphics[height=5.6cm,keepaspectratio]{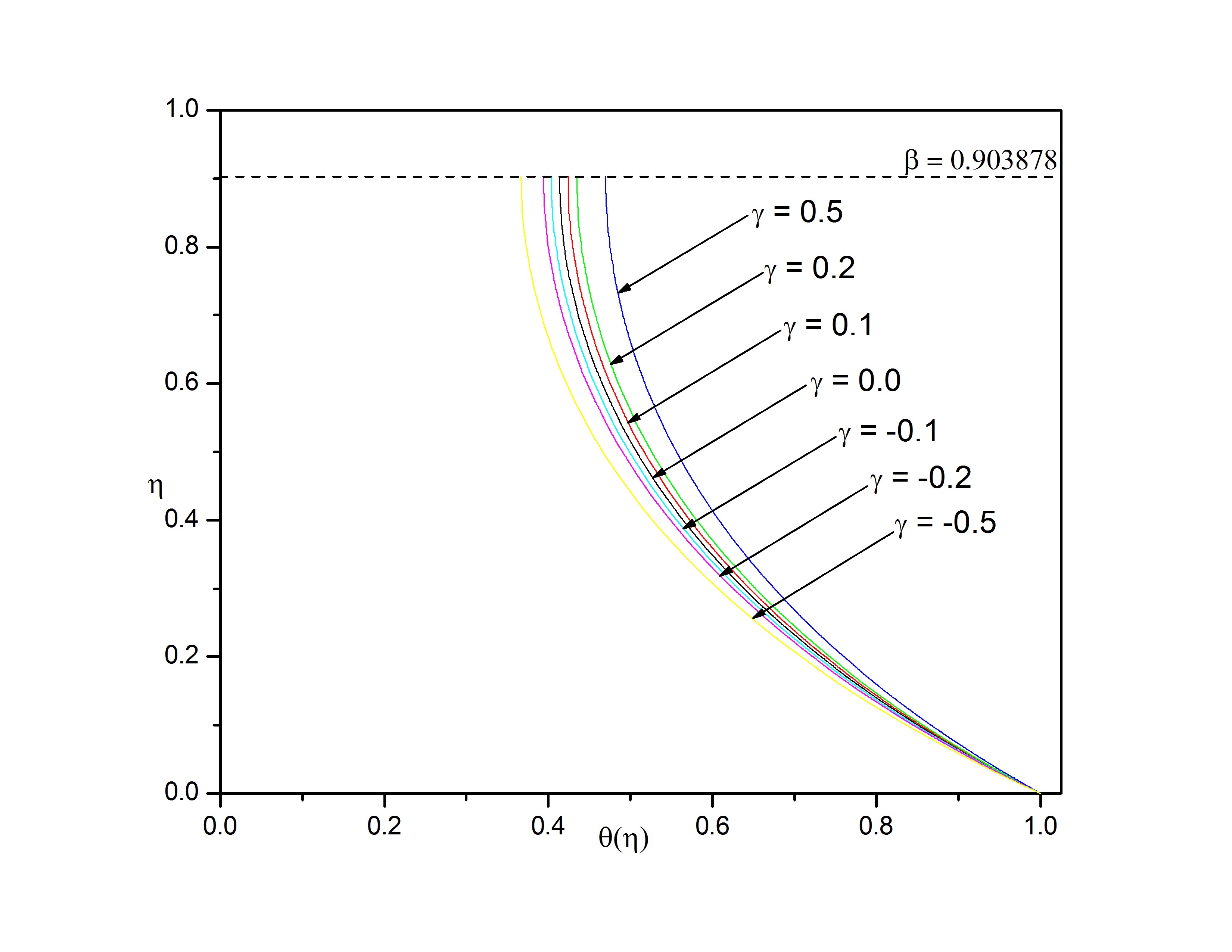}}
\caption{Variation in the temperature profile $ \theta(\eta) $ for different values of Radiation parameter $\gamma$}
 \label{fig7}
\end{figure}

\begin{figure}[htbp]
  \centering
\subfigure[]{\includegraphics[height=5.35cm,keepaspectratio]{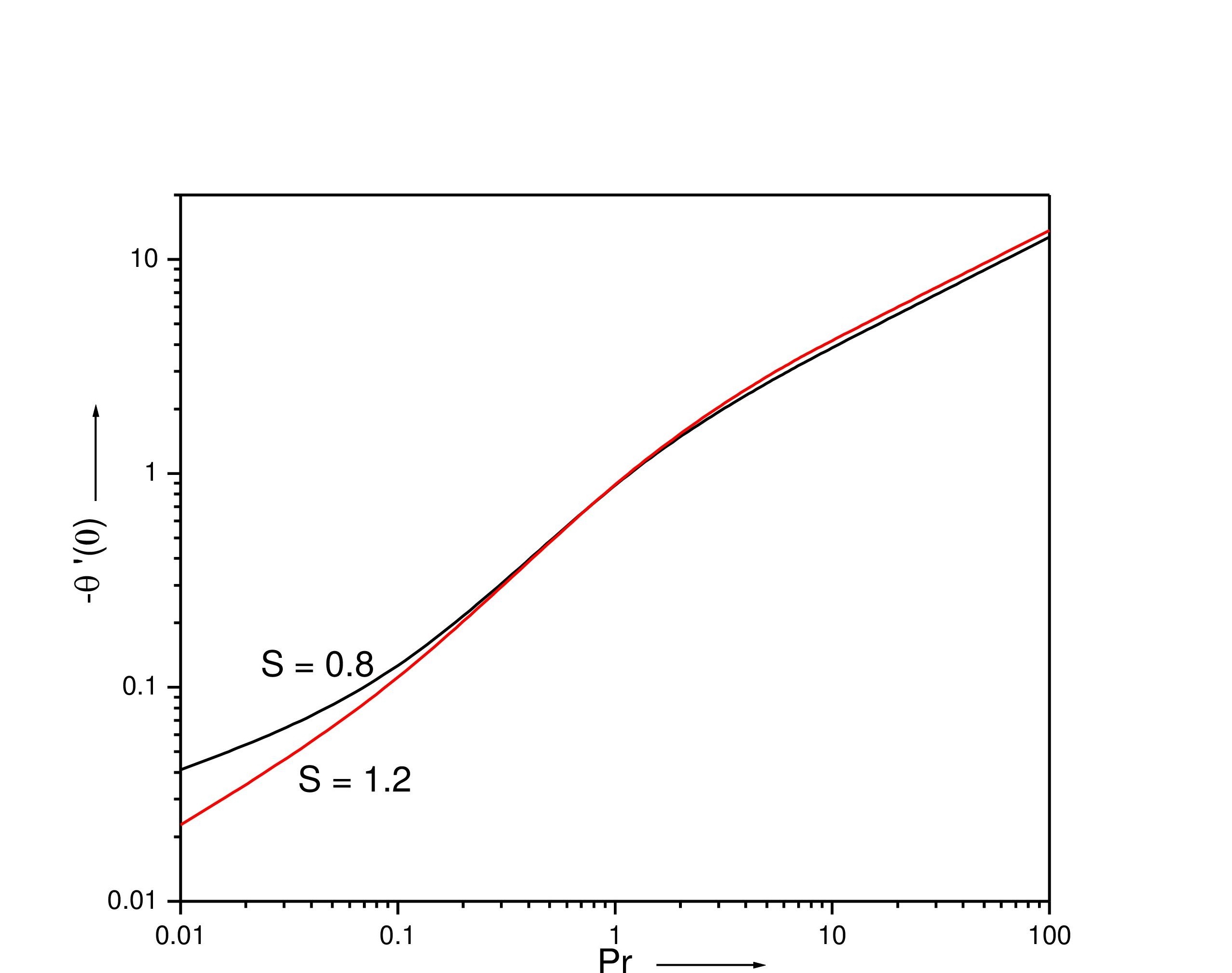}}
\subfigure[]{\includegraphics[height=5cm,keepaspectratio]{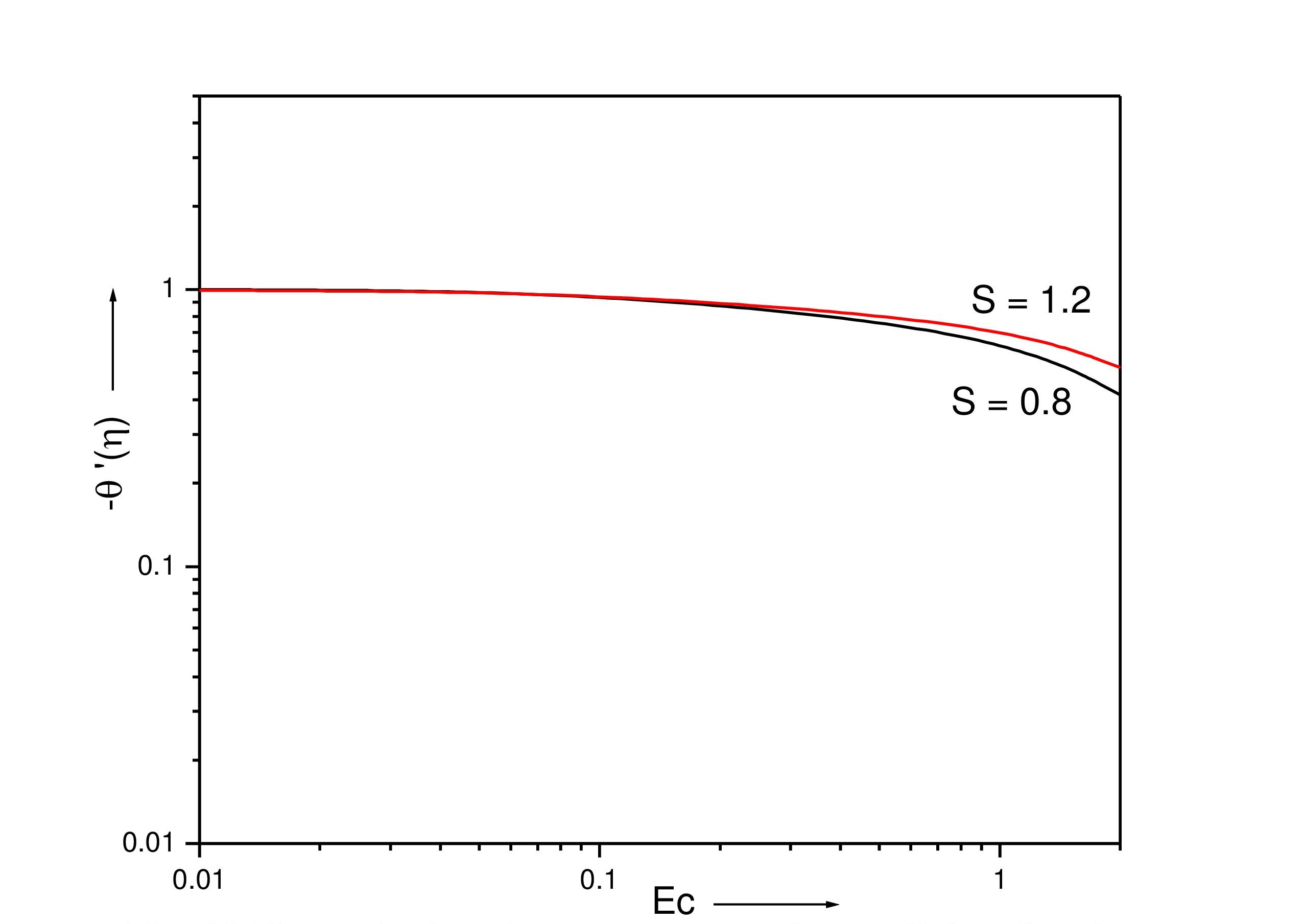}}
\caption{Diemensionless temperature gradiant- $-\theta'(\eta)$ at the sheet vs (a) Prandtl number $Pr$ and (b) Eckert number $Ec$, for S=0.8 and S=1.2}
  \label{fig8}
\end{figure}

\begin{figure}[htbp]
\centering
\includegraphics[width=.55\textwidth]{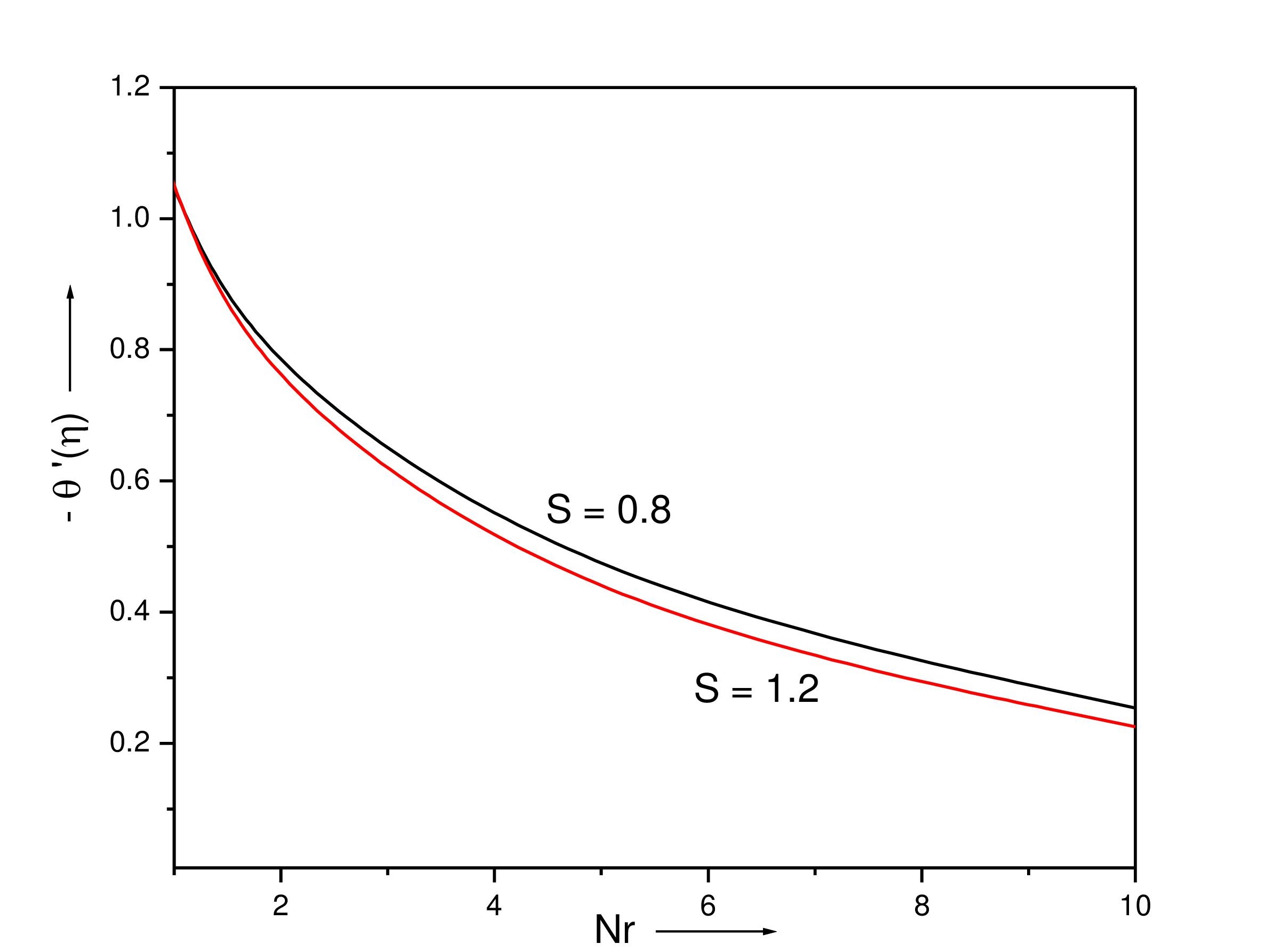}
\caption{Variation of free surface temperature $\theta(\beta)$ with magnetic parameter $Mn$}
\label{fig9}       
\end{figure}

In this work we analyze hydromagnetic thin film flow   and heat transfer characteristics over an unsteady stretching sheet  is investigated, in presence of viscous dissipation, thermal radiation and internal heating. Both numerical and analytical solutions are presented. The similarity transformations were used to transform the governing partial differential equations of flow, heat transfer into a system of non-linear ordinary differential equations. The accuracy of the method was established. The numerical solution obtained  by shooting method together with Runge-Kutta-Fehlberg and Newton-Raphson schemes. It is note worthy to mention that the solution exists only for small value of unsteadiness parameter $ 0\leq S\leq 2 $.  Moreover, when $ S\rightarrow0 $ the solution approaches to the analytical solution obtained by Crane~\cite{3} with infinitely thick layer of fluid $(\beta\rightarrow\infty )$. The other limiting solution corresponding to  $ S\rightarrow2 $ represents a liquid film of infinitesimal thickness $ (\beta\rightarrow 0 )$. The numerical results are obtained for  $ 0\leq S\leq 2 $.  Present results are compared with some of the earlier published results in some limiting cases which are tabulated in Table~\ref{tabtwo}.  The effects of magnetic parameter on various fluid dynamic quantities are shown in Figs.~\ref{fig2}~-~\ref{fig9} for different unsteadiness parameter.\\

Figure~\ref{fig2} highlights the effect of $Mn$ on the dimensionless wall heat flux $ -\theta'(0) $. It is found from this plot that the dimensionless wall heat flux $ -\theta'(0) $  decreases with the increasing values of $Mn$.  The effect of $Mn$ on  $-\theta'(0) $ is observed for different values of unsteadiness parameter $S$.\\

The variation of temperature distribution with magnetic field parameter $Mn$ in  Figures~\ref{fig3}(a) and (b),  temperature distribution for two different values of unsteadiness parameter $S$.  The thermal boundary layer thickness increases with the increasing values of $Mn$.  The increasing frictional drag due to the Lorentz force is responsible for increasing the thermal boundary layer thickness. The results thicken the thermal boundary layer.\\
 
 Figures~\ref{fig4}(a) and (b) demonstrate the effect of Prandtl number $Pr$ on the temperature profiles for two different values of unsteadiness parameter $S$.  These plots reveals the fact that for a particular value of $Pr$ the temperature increases monotonically from the free surface temperature $ T_s $  to wall velocity the  $ T_0 $ as observed by Anderson et al.~\cite{7}.  The thermal boundary layer thickness decreases drastically for high values of $Pr$ i.e., low thermal diffusivity.\\
 
 Figures~\ref{fig5}(a) and (b) project the effect of Eckert number $Ec$ on the temperature distribution for two different values of unsteadiness parameter $S$.  The effect of viscous dissipation is to enhance the temperature in the thin liquid film. i.e., increasing values of $Ec$ contributes in thickening of thermal boundary layer.  For effective cooling of the sheet a fluid of low viscosity is preferable.\\
 
 The effect of radiation parameter $ Nr $ on the horizontal temperature distribution are shown in Figs.~\ref{fig6}(a) and (b) for two different values of unsteadiness parameter $S$. From  these plots it is clear that thermal radiation enhance the temperature in the boundary layer region in the fluid film. Which leads to fall in the rate of cooling for thin film flow. Thermal radiation  defines the rlative contribution of conduction heat transfer to thermal radiation transfer.\\ 
 
Figures \ref{fig7} (a) and (b) present the effect of heat source/sink parameter  $\gamma$ on the temperature distribution for different values of unsteadiness parameter $S$. The effect of sink parameter $\gamma<0$  reduces the temperature in the fluid as the effect of source parameter $\gamma>0$  enhances the temperature. For effective cooling of the sheet, heat sink is preferred. \\
 
 Figures~\ref{fig8}~-~\ref{fig9} present the effect of wall temperature gradient $-\theta'(0)$ at the stretching sheet for $Pr, Ec$ and $Nr$ for different values of unsteadiness parameter $S$.\\
 
Table~\ref{tabtwo} give the comparison of present results with that of Wang~\cite{9}.  With out any doubt, from these tables, we can claim that our results are in excellent agreement with that of Wang~\cite{9} under some limiting cases. \\

\begin{table}
\caption{\label{tabtwo}Comparison of values of surface temperature $\theta(\beta)$ and wall temperature gradiant $-\theta^{'}(0)$ with $Mn=Ec=Nr=A^*=B^*=0.0$.} 
\begin{indented}
\lineup
\item[]\begin{tabular}{@{}*{6}{l}}
\br 
& & $Wang~\cite{9}$ &  \m &$Present ~results$                             &\cr
\br
$\0\0Pr$&$\theta(\beta)$&$-\theta'(0)$&$\frac{\theta'(0)}{\beta}$&$\theta(\beta)$&$-\theta'(0)$\cr 
\mr
& & $S=0.8$ &  \m &$\beta=2.15199$                             &\cr
\mr
\0\00.01&0.960480  &0.090474&$0.042042$&$0.960438$&\00.042120\cr
\0\00.1&0.692533  &0.756162&$0.351378$&$0.692296$&\00.351920\cr
\0\01.0&0.097884  &3.595790&$1.670913$&$0.097825$&\01.671919\cr
\0\02.0&0.024941  &5.244150&$2.436884$&$0.024869$&\02.443914\cr
\0\03.0& 0.008785  &6.514440&$3.027170$&$0.008324$&\03.034915\cr
\mr
& & $S=1.2$ &  \m &$\beta=1.127780$                             &\cr
\mr
\0\00.01&0.982331  &0.037734&$0.033458$&$0.982312$&\00.033515\cr
\0\00.1&0.843622  & 0.343931&$0.304962$&$0.843485$&\00.305409\cr
\0\01.0&0.286717  &1.999590&$1.773032$&$0.286634$&\01.773772\cr
\0\02.0&0.128124  &2.975450 &$2.638324$&$0.128174$&\02.638431\cr
\0\03.0& 0.067658  &3.698830&$3.279744$&$0.067737$&\03.280329\cr
\br
\end{tabular}
\end{indented}
\end{table}

\newpage
\section{Conclusions}
The problem of laminar thin film flow and heat transfer over an unsteady stretching sheet  in presence of variable transverse magnetic field including Viscous dissipation, thermal radiation and internal heating is analyzed. The efficient numerical method of Runge-Kutta-Fehlberg and Newton-Raphson schemes based on shooting technique is used to solve the governing equations. The effects of several parameters controlling the temperature distribution are shown graphically. Some of the important findings of our analysis obtained by the graphical representation are listed below:

\begin{enumerate}
\item	The effect of  Magnetic field on a  electrically conducting fluid is to increase in the thermal boundary layer, by increasing in magnetic parameter $Mn$ the rate of transport of fluid particles reduces due to Lorentz force are generated in the fluid particles, which in turn causes the enhancement of the temperature field.
\item	For a wide range of $  Pr$, the effect Viscous dissipation is found to increase the dimensionless free-surface temperature $ \theta(\beta) $  for the fluid cooling case. The impact of Viscous dissipation on $ \theta(\beta) $  diminishes in the two limiting cases: $ Pr\rightarrow 0, Pr\rightarrow \infty $, in which situations  $ \theta(\beta) $ approaches unity and zero respectively.
\item The energy dissipation (being indicated by the Eckert number) due to heating, viscous dissipation and deformation work has the effect to thicken the thermal boundary layer increases in the temperature profile, and hence reduce the heat transfer rate from the surface. 
\item 	The effect of internal heat source/sink is to generate temperature for increasing positive values and absorb temperature for decreasing  values. However negative value of temperature dependent parameter better suited for cooling purpose.
\item The effect of thermal radiation parameter $Nr$ increases in thermal boundary layer, which in turn causes the enhancement of heat transfer, physically increasing the radiation parameter produces significant increase in the thickness of thermal boundary layer.

\end{enumerate}


\section*{Reference}

\numrefs{1}

\bibitem{1}	Bird, R. B.,  Armstrong, R. C.,  Hassager, O., 1987 Dynamics of polymeric liquids. John Wiley and Sons. New York. \textbf{1}, 

\bibitem{2} Sakiadis, B. C., 1961 Boundary layer behaviour on continuous solid surface: I  -  Boundary layer equations for two dimensional and axisymmetric flow. AIChE. J. \textbf{7},  26--28 

\bibitem{3}	Crane, L. J., 1970 Flow past a stretching plate. Z. Angrew. Math. Phys. \textbf{21}, 645--647 

\bibitem{4}	N. M. Bujurke., S.N.Biradar.,  P.S.Hiremath., 1987 Second order fluid flow past a stretching sheet with heat transfer. ZAMP. \textbf{38}, 653--657

\bibitem{5}	P.C.Ray., B.S.Dandpat., 1994 Flow of thin liquid film on a rotating disk in the presence of a transverse magnetic field. Q. J. Mech. appl. Math.  \textbf{47},  2 

\bibitem{6}	 B. S. Dandpat., P.C.Ray., 1994 The effect of thermocapillarity on the flow of a thin liquid flim on rotating disc. J. Phy: Appl.Phy. \textbf{27},  2041--2045 

\bibitem{7}	Helge. I. Anderson.,  Jan. B. Aarseth.,  Bhabani. S. Dandpat., 2000 Heat transfer in a liquid film on an unsteady stretching stretching surface. Int. J. H. and Mass Tr. \textbf{43},  69--74 

\bibitem{8}	Bhabani. S. Dandpat.,  Bidyut. Santra., 2003 Thermocapillarity in a liquid film on an unsteady stretching surface. J. H. and Mas. Tra.  \textbf{46}, 3009--3015 

\bibitem{9}	Wang, C. 2006 Analytic solutions for a liquid film on an unsteady stretching surface. H. M. Tra. \textbf{42},  759-–766 

\bibitem{10}	B. S. Dandpat., S. Maity., 2006 Flow of a thin liquid film on an unsteady stretching sheet. Phy. of flu. \textbf{18},102101  

\bibitem{11}	Bidyut. Santra.,  Bhabani. S. Dandpat.,  2009 Unsteady thin fil flow over a heated stretching sheet. Int. J. H. and M. Tra. \textbf{52}, 1965--1970 

\bibitem{12} N. F. M. Noor., I. Hashim., 2010  Thermocapillary and magnetic field effects in a thin liquid film on an unsteady stretching surface. Int. J. H. and M. Tra. \textbf{53}, 2044--2051 

\bibitem{13}	B. S. Dandpat., S. Chakraborthy., 2010 Effects of variable fluid propeties on unsteady thin film flow over a non-linear stretching sheet. Int. J. H. and M. Tra.  \textbf{53}, 5757--5763 

\bibitem{14}	B. S. Dandpat., S. K. Singh., 2011 Two layer film flow on ratating disc:A numerical study. Int. J. of Non-lin. mech.  \textbf{46},272--277 

\bibitem{15} B. S. Dandpat., S. K. Singh., 2011 Thin film flow over a heated nonlinear stretching sheet in presence off uniform transverse magnetic field. Int. J. H. and M. Tra. \textbf{38}, 324--328


\bibitem{16}	I-Chung. Liu., Ahmed. M. Megahed., 2012 Homotopy petrubution method for thin film flow and heat transfer over an unsteady stretching sheet with internal heating variable heat flux. J. of Appl. Math. Id:418527 doi:10.1155/2012/418527 

\bibitem{17}	K. Vajravelu., K. V. Prasad., Chiu-On. Ng., 2012 Unsteady flow and heat transfer in a thin film of Ostwald-de Waele liquid over a stretching surface. Comm.nonlin. Sci. Num. Sim. \textbf{17},  4163-–4173 

\bibitem{18}	R. C. Aziz., I. Hashim., A. K. Alomari., 2013 Flow and heat transfer in a liquid film over a permeable stretching sheet. J. appl. Math. ID:487586, doi:10.1155/2013/487586

\bibitem{19} Taza. Gul., Saeed. Islam., Rehan. Ali. Shah., Ilyas. Khan., Asma. Khalid., Sharidan. Shaifie 2014  Heat transfer analysys of mhd thin film flow of an unsteady second grade fluid past a vertical oscillating belt. Plos one \textbf{9(1)},  1--21 

\bibitem{20} A. M. Megahed., 2015 Effect of slip velocity on casson thin film flow and heat transfer due to stretching sheet in presence of variable heat flux and viscous dissipation. Appl. Math. Mech. Engl. Ed ,  DOI .10.1007/s10483-015-1983-9.

\bibitem{21} Alice. B. Thompson., Susana. N. Gomes., Grigorios. A. Pavliotis., Demetrios. T. papgeorgiou., 2016 Stabilising falling liquid film flows using feedback control, \textbf{28}, DOI.http://dx.doi.org/10.1063/1.4938761


\bibitem{22}	Conte, S. D.,  De Boor, C., 1972 Elementary Numerical Analysis. McGraw-Hill. New York. 

\bibitem{23}	Brewster, M. Q.,  1972 Thermal radiative transfer properties. John wileys and sons.
\endnumrefs
\end{document}